\documentclass[12pt]{article}
\usepackage{graphicx}
\usepackage{amssymb}

\begin{document}
\title{
Deconstructing the Gel'fand--Yaglom method and vacuum energy from a theory
space}
\author{
Nahomi Kan\\
{\footnotesize 
National Institute of Technology, Gifu College,
Motosu-shi, Gifu 501-0495, Japan}\\
{\small and}\\
Kiyoshi Shiraishi\\
{\footnotesize Graduate School of Sciences and Technology for Innovation,}
\\
{\footnotesize  Yamaguchi University, Yamaguchi-shi, Yamaguchi 753--8512,
Japan} }
\date{\today}
\maketitle

\abstract{
The discrete Gel'fand--Yaglom theorem was 
studied 
several years ago.
In the present paper, we generalize the discrete Gel'fand--Yaglom
method to obtain the determinants of mass matrices which appear
current works in particle physics, such as dimensional deconstruction
and clockwork theory. Using the results, we show the expressions for
vacuum energies in such various models. 
}


\section{Introduction}

The Gel'fand--Yaglom method \cite{GY} for obtaining functional
determinants of differential operators with boundaries is widely known
nowadays. For nice reviews, see
\cite{Dunne,Coleman}. The applications of the Gel'fand--Yaglom method have
been investigated quite recently, to evaluate one-loop vacuum energies
in nontrivial boundary conditions
\cite{TFM,FM,Altshuler1,Altshuler2}.

Among them, Altshuler examined vacuum energy in warped compactification
\cite{Altshuler1,Altshuler2}. In recent
years, it is supposed that extra dimensions of various types could play an
important role in the hierarchy problem, and thus the study of
physics in nontrivial background geometry is still advancing.

The dimensional deconstruction has appeared as a new tool for
understanding the properties of higher-dimensional field theories
\cite{ACG,HPW,HL} more than a decade ago. In such a model of
deconstruction,  a `theory space' is considered, which consists of sites
and links, to which four-dimensional fields are individually assigned.
Theory spaces thus have the structures of graphs \cite{Wilson}
and can be interpreted as the theory with discrete extra dimensions.

Several years ago, the discrete Gel'fand--Yaglom method for difference
operators was reviewed and studied by Dowker \cite{Dowker}.
We generalize the discrete Gel'fand--Yaglom method for studying 
one-loop vacuum energies in extended deconstructed theories and models
with discrete dimensions in the present paper.
To this ends, we develop the method of computing determinants of
repetitive Hermitian matrices which correspond to mass matrices
utilized in deconstructed theories.

After completion of the first version of the manuscript of the present
paper ({\tt arXiv:1711.06806}), a paper which treats the determinants of
discrete Laplace operators appeared \cite{Ossipov}. Their method is
substantially the same as ours, because the author also relies on the
recurrence relation among three variables on a lattice (see
Sec.~\ref{sec3} in the present paper and below). We recently become aware
of another similar paper on the determinants of matrix differential
operators
\cite{FFG}.   They studied generalization of Gel'fand--Yaglom method to
obtain the functional determinants. Their work differs essentially from
ours because they considered differential operators while we treat
matrices as operators. We also point out that they did not consider the
matrices of large size which have certain continuum limits.

The organization of this paper is as follows.
In order to make the present paper self-contained, we show a short review
of the Gel'fand--Yaglom method for a differential operator, along
with the Dunne's review \cite{Dunne}, in Sec.~\ref{sec2}. 
In Sec.~\ref{sec3}, we give the method to obtain determinants of
tridiagonal matrices with repeated structure. This is a
straightforward generalization of description in Ref.~\cite{Dowker}.
In Sec.~\ref{sec4}, we give the method to obtain determinants of
periodic tridiagonal matrices. Determinants of extended periodic
tridiagonal matrices are obtained in Sec.~\ref{sec5}.
The rest of the present paper is devoted to applications to
deconstructed theories and discrete systems.
In Sec.~\ref{sec6}, free energy on a graph is discussed by using the
results of previous sections.
In Sec.~\ref{sec7}, we show the method of calculation for
evaluating one-loop vacuum energy in deconstructed models from the
determinants of mass matrices. In Sec.~\ref{sec8}, we show a few more
examples of one-loop vacuum energies for slightly complicated theory
spaces. We give conclusions in the last section, Sec.~\ref{concl}.

\section{Review of the Gel'fand--Yaglom method \cite{Dunne}}
\label{sec2}

Suppose that an eigenvalue equation
$(\Delta+m^2)\psi^{(m^2)}_{\lambda}(x)=\lambda^{(m^2)}\psi^{(m^2)}_
{\lambda}(x)$ with Dirichlet-Dirichlet boundary conditions
$\psi^{(m^2)}_{\lambda}(0)=\psi^{(m^2)}_{\lambda}(L)=0$ in one dimension
is given.

Then
\begin{equation}
\frac{\det[\Delta+m^2]}{\det[\Delta]}=\frac{\psi^{(m^2)}_{0}(L)}
{\psi^{(0)}_{0}(L)}\,,
\end{equation}
holds.
Here
$\psi^{(m^2)}_{0}(x)$ satisfies
$(\Delta+m^2)\psi^{(m^2)}_{0}(x)=0$ with a boundary condition
$\psi^{(m^2)}_{0}(0)=0$.

\subsubsection*{example}
In the region $0\le x\le L$, under the Dirichlet condition at the
boundaries, we consider the functional determinant
\begin{equation}
\frac{\det[-\partial_x^2+m^2]}{\det[-\partial_x^2]}\,.
\end{equation}
The solution of $(-\partial_x^2+m^2)\psi(x)=0$ with
$\psi(0)=0$ is 
$\sinh mx$.
Thus according to the
Gel'fand--Yaglom method, we obtain
\begin{equation}
\frac{\det[-\partial_x^2+m^2]}{\det[-\partial_x^2]}=\frac{\sinh mL}{mL}
=\prod_{n=1}^\infty\frac{\frac{n^2\pi^2}{L^2}+m^2}{\frac{n^2\pi^2}{L^2}}\,.
\end{equation}

\subsubsection*{proof}
$\det[\Delta+m^2-\lambda]$ is a function of $\lambda$ and has zeros at
$\lambda=\lambda^{(m^2)}$. The function $\psi_\lambda(x)$ which satisfies
$(\Delta+m^2-\lambda)\psi_\lambda(x)=0$ and $\psi_\lambda(0)=0$ becomes
the eigenfunction when
$\lambda=\lambda^{(m^2)}$. Then the boundary condition at $x=L$, i.e.,
$\psi_\lambda(L)=0$ is satisfied. In other words,
$\psi_\lambda(L)$ is a function of $\lambda$ and has zeros at
$\lambda=\lambda^{(m^2)}$. Therefore
$\det[\Delta+m^2-\lambda]\propto\psi_\lambda(L)$ holds.\rule{5pt}{10pt}

\section{Determinants of tridiagonal matrices}
\label{sec3}

\subsection{the discrete Gel'fand--Yaglom method for tridiagonal matrices}

Now, we show the disctrete Gel'fand--Yaglom method to obtain
determinants of finite matrices.
First, we consider the following Hermitian tridiagonal
matrix of
$N$ rows and columns:
\begin{equation}
H=\left(
\begin{array}{cccccc}
c&-b&0&\cdots& &0\\
-b^*&a&-b&\cdots& &0\\
0&-b^*&a&\cdots& &0\\
\vdots&\vdots&\vdots&\ddots& &\vdots\\
 & & & &a&-b\\
0&0&0&\cdots&-b^*&d
\end{array}
\right)\,.
\end{equation}
In this case, the eigenvalue equation
\begin{equation}
H\mathbf{v}=\lambda\mathbf{v}\,,
\end{equation}
where $\mathbf{v}^T=(v_1,v_2,\dots,v_N)$,
can be categorized into three parts,
\begin{equation}
\sum_{j=1}^NH_{1j}v_j-\lambda v_1=0\,,
\label{ini}
\end{equation}
\begin{equation}
\sum_{j=1}^NH_{kj}v_j-\lambda v_k=0\quad (2\le k\le N-1)\,,
\label{rr}
\end{equation}
\begin{equation}
\sum_{j=1}^NH_{Nj}v_j-\lambda v_N=0\,.
\label{de}
\end{equation}
Here, Eq.~(\ref{rr}) is just the recurrence relation among
three terms in
$v_k$ as a sequence of numbers.
In the present case, the general solution for the recurrence relation
\begin{equation}
bv_{k+1}-(a-\lambda)v_k+b^*v_{k-1}=0\quad (2\le k\le N-1)
\end{equation}
is
\begin{equation}
v_k=A\alpha^{k-1}+B\beta^{k-1}\,,
\end{equation}
where $A$ and $B$ are constants and
\begin{equation}
\alpha=\frac{a-\lambda+\sqrt{(a-\lambda)^2-4|b|^2}}{2b}\,,
\quad\beta=\frac{a-\lambda-\sqrt{(a-\lambda)^2-4|b|^2}}{2b}\,.
\label{alphabeta}
\end{equation}
Note that $\alpha$ and $\beta$ are roots of the second-order equation
$bx^2-(a-\lambda)x+b^*=0$ and $\alpha\beta=b^*/b$.

The first row of the eigenvalue equation, Eq.~(\ref{ini}), determines the
relation between
$v_1$ and
$v_2$; in this case, that is $(c-\lambda)v_1-bv_2=0$. If we further choose
\begin{equation}
v_1=A+B=1\,,
\end{equation}
the coefficients $A$ and $B$ are obtained as
\begin{equation}
A=A(\lambda)=\frac{c-\lambda-b\beta}{b(\alpha-\beta)}
=\frac{c-\lambda-b\beta}{\sqrt{(a-\lambda)^2-4|b|^2}}\,,
\end{equation}
\begin{equation}
B=B(\lambda)=\frac{-(c-\lambda-b\alpha)}{b(\alpha-\beta)}
=\frac{-(c-\lambda-b\alpha)}{\sqrt{(a-\lambda)^2-4|b|^2}}\,.
\end{equation}

Substituting all of the results above into Eq.~(\ref{de}) in the present
case, we get
\begin{equation}
A(\lambda)[-b^*+(d-\lambda)\alpha]\alpha^{N-2}+
B(\lambda)[-b^*+(d-\lambda)\beta]\beta^{N-2}=0\,.
\label{ld}
\end{equation}

Now, we set the left-hand side of Eq.~(\ref{ld}) as $\tilde{D}(\lambda)$.
$\tilde{D}(\lambda)$ is zero if $\lambda$ is an eigenvalue of the matrix
$H$ in this case.
By construction, $\tilde{D}(\lambda)$ should be an $N$th order polynomial
of $\lambda$. The reason is: $v_2=[(c-\lambda)/b]v_1=(c-\lambda)/b$,
$v_3=[(a-\lambda)/b]v_2-(b^*/b)v_1=(a-\lambda)(c-\lambda)/b^2-(b^*/b)$,
and so on. This observation shows $v_{N}$ includes
$(-{\lambda})^{N-1}/b^{N-1}$. Finally, since the left-hand side of
Eq.~(\ref{de}) reads
$-b^*v_{N-1}+(d-\lambda)v_N$ in the present case, $\tilde{D}(\lambda)$
has the term $(-{\lambda})^{N}/b^{N-1}$ as the highest order term in
$\lambda$. We can also directly confirm this by setting
$a=c=d=0$ and the limit
$b\rightarrow 0$ in the left-hand side of Eq.~(\ref{ld}). We then verify
$\tilde{D}(\lambda)\rightarrow(-{\lambda})^N/b^{N-1}$.

Therefore, we conclude that
$D(\lambda)=b^{N-1}\tilde{D}(\lambda)=\prod_{p=1}^N(\lambda_p-\lambda)$
is the characteristic polynomial of $H$,
where $\lambda_p$ $(p=1, 2, \dots, N)$ are eigenvalues of $H$.

The determinant of $H$ is given by $D(0)=\prod_{p=1}^N\lambda_p$. In the
present case, we find
\begin{equation}
\det
H=D(0)=\frac{b^{N-1}}{\sqrt{a^2-4|b|^2}}\Big[(c-b\bar{\beta})(-b^*+d\bar\alpha)
\bar\alpha^{N-2}-(c-b\bar\alpha)(-b^*+d\bar\beta)\bar\beta^{N-2}\Big]\,,
\end{equation}
where
\begin{equation}
\bar\alpha=\frac{a+\sqrt{a^2-4|b|^2}}{2b}\,,
\quad\bar\beta=\frac{a-\sqrt{a^2-4|b|^2}}{2b}\,.
\end{equation}
After a lengthy calculation, we obtain
\begin{eqnarray}
& &\det H=D(0)\nonumber \\
&=&\frac{1}{2^N\sqrt{a^2-4|b|^2}}\Big\{\Big[2(cd-|b|^2)+
(a-c-d)(a-\sqrt{a^2-4|b|^2})\Big]
(a+\sqrt{a^2-4|b|^2})^{N-1}\nonumber \\
& &-\Big[2(cd-|b|^2)+(a-c-d)(a+\sqrt{a^2-4|b|^2})\Big]
(a-\sqrt{a^2-4|b|^2})^{N-1}\Big\}\,\nonumber \\& &
\qquad\mbox{(for a
tridiagonal matrix)}\,,
\label{trid}
\end{eqnarray}
in the present case.
It is notable that the determinant depends only on $|b|$ and
does not depend on $b^*/b$ in the present case.
The reason is because the eigenvalues are unchanged under ``gauge''
transformation
$\mathbf{v}\rightarrow P\mathbf{v}$ and $H\rightarrow PHP^{-1}$,
where $P=diag.(1, e^{i\chi},  e^{i2\chi},\ldots,  e^{i(N-1)\chi})$
with an arbitrary real constant $\chi$.

The prescription of the above method to obtain the determinant is very
similar to the Gel'fand--Yaglom method for differential operators.
Namely, solving the differential equation corresponds to solving the
recurrence relation, putting one of the boundary conditions corresponds
to fixing the first term of the series of numbers, and obtaining the
determinant at another boundary corresponds to obtaining the determinant
as the equation of the last row in the eigenvalue equation. 
Note that, because we are treating a finite matrix, the idea of
normalization becomes different from the functional determinant
treated by the Gel'fand--Yaglom method.

The method to obtain the determinant of tridiagonal matrices in this
section is substantially equivalent to  the method for difference
operators described by Dowker
\cite{Dowker}, except for a specific choice for an Hermitian matrix in the
present section.

\subsection{examples}

In this subsection, we show determinants of some simple
tridiagonal matrices for example.
For all the examples below, the eigenvalues are known
and then, one can find that the formulas%
\footnote{For example, see \cite{ZJ}.}
 for finite product including
trigonometric functions are derived.

Note that the determinant $D(0)$ for the matrix $H=\Delta+m^2 I$ (where
$I$ is the identity matrix) is equivalent to
$D(-m^2)$ for the matrix
$\Delta$, we choose explicit expressions of $D(0)$ for $H$ here and
hereafter.

\subsubsection*{\textbullet~ $a=c=d=2+4\sinh^2\frac{z}{2}$ and $b=1$.%
\footnote{In this simple case, eigenvalues are known as
\begin{equation}
\lambda_p=4\sin^2\frac{\pi p}{2(N+1)}+4\sinh^2\frac{z}{2}\quad
(p=1, 2, \ldots, N)\,.
\end{equation}}}

In this case, $H=\Delta_{DD}+4\sinh^2\frac{z}{2}\,I_N$,
where
\begin{equation}
\Delta_{DD}\equiv\left(
\begin{array}{cccccc}
2&-1&0&\cdots& &0\\
-1&2&-1&\cdots& &0\\
0&-1&2&\cdots& &0\\
\vdots&\vdots&\vdots&\ddots& &\vdots\\
&&&&2&-1\\
0&0&0&\cdots&-1&2
\end{array}
\right)\,,
\end{equation}
and $I_N$ is the $N\times N$ identity matrix.
Noting that, for $a=c=d$, Eq.~(\ref{trid}) becomes
\begin{equation}
D(0)=\frac{1}{2^{N+1}\sqrt{a^2-4|b|^2}}\Big[
(a+\sqrt{a^2-4|b|^2})^{N+1}-
(a-\sqrt{a^2-4|b|^2})^{N+1}\Big]\,.
\end{equation}
We now find
\begin{equation}
\det H=D(0)=\frac{\sinh (N+1)z}{\sinh z}\quad
\mbox{for~} H=\Delta_{DD}+4\sinh^2\frac{z}{2}\,I_N\,.
\label{dDD}
\end{equation}

\subsubsection*{\textbullet~ $a=c=2+4\sinh^2\frac{z}{2}$,
$d=1+4\sinh^2\frac{z}{2}$ and $b=1$.%
\footnote{In this case,
\begin{equation}
\lambda_p=4\sin^2\frac{\pi (p-1/2)}{2(N+1/2)}
+4\sinh^2\frac{z}{2}\quad
(p=1, 2, \ldots, N)\,.
\end{equation}
}}

In this case, $H=\Delta_{DN}+4\sinh^2\frac{z}{2}\,I_N$,
where
\begin{equation}
\Delta_{DN}\equiv\left(
\begin{array}{cccccc}
2&-1&0&\cdots& &0\\
-1&2&-1&\cdots& &0\\
0&-1&2&\cdots& &0\\
\vdots&\vdots&\vdots&\ddots& &\vdots\\
&&&&2&-1\\
0&0&0&\cdots&-1&1
\end{array}
\right)
\end{equation}
We find
\begin{eqnarray}
\det H=D(0)&=&\frac{\sinh (N+1)z-\sinh Nz}{\sinh z}\nonumber \\
&=&\cosh Nz+2\sinh z\sinh Nz\quad
\mbox{for~} H=\Delta_{DN}+4\sinh^2\frac{z}{2}\,I_N\,.
\label{dDN}
\end{eqnarray}

\subsubsection*{\textbullet~ $a=2+4\sinh^2\frac{z}{2}$,
$c=d=1+4\sinh^2\frac{z}{2}$ and $b=1$.%
\footnote{In this case, the eigenvalues are
\begin{equation}
\lambda_p=4\sin^2\frac{\pi p}{2N}+4\sinh^2\frac{z}{2}\quad
(p=0, 1, 2, \ldots, N-1)\,.
\end{equation}
}
}

\begin{figure}[ht]
\centering
\includegraphics[height=5cm]
{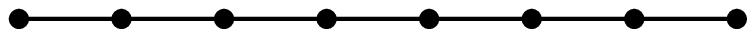}
\caption{%
$P_{8}$ as an example of a path graph.
}
\label{path}
\end{figure}

In this case, $H=\Delta_{NN}+4\sinh^2\frac{z}{2}\, I_N$,
where
\begin{equation}
\Delta_{NN}\equiv\left(
\begin{array}{cccccc}
1&-1&0&\cdots& &0\\
-1&2&-1&\cdots& &0\\
0&-1&2&\cdots& &0\\
\vdots&\vdots&\vdots&\ddots& &\vdots\\
&&&&2&-1\\
0&0&0&\cdots&-1&1
\end{array}
\right)=\Delta(P_N)\,,
\end{equation}
which is known as the graph Laplacian \cite{Mohar1,Mohar2,Mohar3,Merris}
for the path graph with $N$ vertices ($P_N$, see FIG.~\ref{path}).

We find
\begin{equation}
\det H=D(0)=2\tanh\frac{z}{2}\sinh Nz\quad
\mbox{for~} H=\Delta_{NN}+4\sinh^2\frac{z}{2}\,I_N\,
\label{dNN}
\end{equation}
in this case.
Note that, since $\Delta(P_N)$ has a zero mode,
$\lim_{z\rightarrow 0}D(0)=0$.

\subsubsection*{\textbullet~ clockwork theory \cite{CI,KR,GM,FPRT}.%
\footnote{In this case, 
\begin{equation}
\lambda_0=l^2\,,\quad\lambda_p=q^2+1-2q\cos\frac{\pi
p}{N}+l^2\quad (p=1, 2, \ldots, N-1)\,.
\end{equation}
}
}

We consider $H=\Delta_q+l^2I_N$, where $\Delta_q$ is the following
$N\times N$ matrix:
\begin{equation}
\Delta_q\equiv\left(
\begin{array}{cccccc}
1&-q&0&\cdots& &0\\
-q&1+q^2&-q&\cdots& &0\\
0&-q&1+q^2&\cdots& &0\\
\vdots&\vdots&\vdots&\ddots& &\vdots\\
&&&&1+q^2&-q\\
0&0&0&\cdots&-q&q^2
\end{array}
\right)\,.
\end{equation}
We find that the determinant of $H$ can be written as
\begin{equation}
D(0)=\frac{l^2\cdot(\gamma_+^N-\gamma_-^N)}{
\sqrt{(1-q^2)^2+2(1+q^2)l^2+l^4}}\quad
\mbox{for~} H=\Delta_q+l^2I_N\,,
\label{cw}
\end{equation}
where 
\begin{equation}
\gamma_\pm=\frac{1}{2}\left[1+q^2+l^2\pm
\sqrt{(1-q^2)^2+2(1+q^2)l^2+l^4}\right]\,.
\label{gam}
\end{equation}
Of course, one can see that $\lim_{q\rightarrow 1}\Delta_q=\Delta(P_N)$.

\section{Determinants of periodic tridiagonal matrices}
\label{sec4}

\subsection{the discrete Gel'fand--Yaglom method for periodic tridiagonal
matrices}

In this section, we treat periodic tridiagonal matrices, such as
\begin{equation}
H=\left(
\begin{array}{cccccc}
a&-b&0&\cdots& &-b^*\\
-b^*&a&-b&\cdots& &0\\
0&-b^*&a&\cdots& &0\\
\vdots&\vdots&\vdots&\ddots& &\vdots\\
 & & & &a&-b\\
-b&0&0&\cdots&-b^*&a
\end{array}
\right)\,.
\end{equation}

In this case,
the recurrence relation is same as in the previous section.
Therefore, we can write
\begin{equation}
v_k=A\alpha^{k-1}+B\beta^{k-1}\,,
\end{equation}
where $\alpha$ and $\beta$ are same as the previous ones, i.e.,
Eq.~(\ref{alphabeta}).

In the periodic case, however, the first and the last rows of the
eigenvalue equation are also the relation among three terms in the
sequence of numbers. In the present case, they are reduced to
\begin{equation}
b[A\beta(1-\alpha^{N})+B\alpha(1-\beta^{N})]=0\,,
\label{c1}
\end{equation}
\begin{equation}
b[A\alpha(\alpha^{N}-1)+B\beta(\beta^{N}-1)]=0\,,
\label{cN}
\end{equation}
where we used the fact that $\alpha$ and $\beta$ are solutions of
$bx^2-(a-\lambda)x+b^*=0$ and $\alpha\beta=b^*/b$.
The existence of $A$ and $B$ satisfying the above two equations
and not being $A=B=0$
requires
\begin{equation}
\left|\begin{array}{cc}
\beta(1-\alpha^{N})&\alpha(1-\beta^{N})\\
\alpha(\alpha^{N}-1)&\beta(\beta^{N}-1)
\end{array}
\right|=(\beta^2-\alpha^2)(1-\alpha^{N})(\beta^{N}-1)=0\,.
\end{equation}
This equation is satisfied if $\lambda$ is an eigenvalue of the matrix
$H$. In general, we suppose $\alpha\ne\beta$ and the normalization can be
known from the limit $a=0$ and $b\rightarrow 0$.
Then, we conclude that the
characteristic polynomial $D(\lambda)=\prod_{p=1}^N(\lambda_p-\lambda)$
(where $\lambda_p$ ($p=1,\dots,N$) are eigenvalues of $H$) is written by
\begin{equation}
D(\lambda)=b^N(1-\alpha^{N})(\beta^{N}-1)=b^N(\alpha^{N}+\beta^{N})
-b^N-{b^*}^N\,.
\end{equation}
Therefore, the determinant of $H$ in this case is given by
\begin{eqnarray}
& &\det H=D(0)=\left(\frac{a+\sqrt{a^2-4|b|^2}}{2}\right)^{N}+
\left(\frac{a-\sqrt{a^2-4|b|^2}}{2}\right)^{N}
-b^N-{b^*}^N\nonumber \\
& &\qquad\mbox{(for a periodic tridiagonal matrix)}\,.
\end{eqnarray}

One may be aware of unnecessary arguments in above discussion.
From the periodic structure, $\alpha^N=1$ or $\beta^N=1$ can be
concluded. However, the discussion above can be generalized to treat
another type of matrix in the next section.  

\subsection{example}

\subsubsection*{\textbullet~ $a=2+4\sinh^2\frac{z}{2}$ and $b=1$.%
\footnote{In this case, the eigenvalues are
\begin{equation}
\lambda_p=4\sin^2\frac{\pi p}{N}+4\sinh^2\frac{z}{2}\quad
(p=0, 1, 2, \ldots, N-1)\,.
\end{equation}
Note that degeneracy occurs.
}
}

\begin{figure}[ht]
\centering
\includegraphics[height=5cm]
{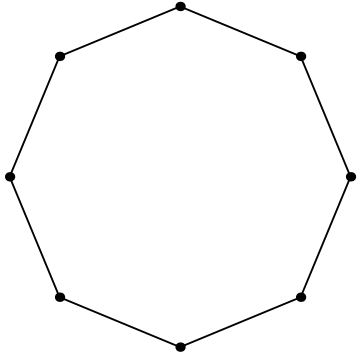}
\caption{%
$C_{8}$ as an example of a cycle graph.
}
\label{cycle}
\end{figure}

In this case, $H=\Delta(C_N)+4\sinh^2\frac{z}{2}\,I_N$,
where $\Delta(C_N)$ is the graph Laplacian of 
the cycle graph with $N$ vertices (see FIG.~\ref{cycle}),
\begin{equation}
\Delta(C_N)\equiv\left(
\begin{array}{cccccc}
2&-1&0&\cdots& &-1\\
-1&2&-1&\cdots& &0\\
0&-1&2&\cdots& &0\\
\vdots&\vdots&\vdots&\ddots& &\vdots\\
 & & & &2&-1\\
-1&0&0&\cdots&-1&2
\end{array}
\right)\,.
\end{equation}

We find
\begin{equation}
\det H=D(0)=4\sinh^2\frac{Nz}{2}=\left(e^{Nz/2}-e^{-Nz/2}\right)^2\quad
\mbox{for~} H=\Delta(C_N)+4\sinh^2\frac{z}{2}\,I_N\,.
\label{dcyc}
\end{equation}
Note that $\lim_{z\rightarrow 0}D(0)=0$ because of the zero mode of
$\Delta(C_N)$.

\subsubsection*{\textbullet~ $a=2+4\sinh^2\frac{z}{2}$ and $b=e^{i\chi}$.%
\footnote{In this case, the eigenvalues are
\begin{equation}
\lambda_p=4\sin^2(\frac{\pi p}{N}+\frac{\chi}{2})+4\sinh^2\frac{z}{2}\quad
(p=0, 1, 2, \ldots, N-1)\,.
\end{equation}}
}

In this case, $H=\Delta(C_N,\chi)+4\sinh^2\frac{z}{2}\,I_N$,
where 
\begin{equation}
\Delta(C_N,\chi)\equiv\left(
\begin{array}{cccccc}
2&-e^{i\chi}&0&\cdots& &-e^{-i\chi}\\
-e^{-i\chi}&2&-e^{i\chi}&\cdots& &0\\
0&-e^{-i\chi}&2&\cdots& &0\\
\vdots&\vdots&\vdots&\ddots& &\vdots\\
 & & & &2&-e^{i\chi}\\
-e^{i\chi}&0&0&\cdots&-e^{-i\chi}&2
\end{array}
\right)\,.
\end{equation}
We find
\begin{eqnarray}
\det H=D(0)&=&4\sinh^2\frac{Nz}{2}+4\sin^2\frac{N\chi}{2}\nonumber \\
&=&\left|e^{N(z+i\chi)/2}-e^{-N(z+i\chi)/2}\right|^2\quad
\mbox{for~} H=\Delta(C_N,\chi)+4\sinh^2\frac{z}{2}\,I_N\,.
\label{dcycchi}
\end{eqnarray}

\section{Determinants of extended periodic tridiagonal matrices}
\label{sec5}

\subsection{the discrete Gel'fand--Yaglom method for extended
periodic tridiagonal matrices}

In this section, we consider the following $(N+1)\times(N+1)$ matrix
\begin{equation}
H=\left(
\begin{array}{ccccccc}
a&-b&0&\cdots& &-b^*&-d\\
-b^*&a&-b&\cdots& &0&-d\\
0&-b^*&a&\cdots& &0&-d\\
\vdots&\vdots&\vdots&\ddots& &\vdots&\vdots\\
 & & & &a&-b&-d\\
-b&0&0&\cdots&-b^*&a&-d \\
-d^*&-d^*&-d^*&\cdots&-d^*&-d^*&c
\end{array}
\right)\,.
\end{equation}
The recurrence relation can be found as
\begin{equation}
bv_{k+1}-(a-\lambda)v_k+b^*v_{k-1}=-d\,v_{N+1}\quad (2\le k\le N-1)\,.
\end{equation}
The general solution of this equation is
\begin{equation}
v_k=A\alpha^{k-1}+B\beta^{k-1}-\frac{d\,v_{N+1}}{b+b^*-(a-\lambda)}\,,
\end{equation}
where $\alpha$ and $\beta$ are same as Eq.~(\ref{alphabeta}).

The first row of the eigenvalue equation then becomes
\begin{equation}
b[A\beta(1-\alpha^{N})+B\alpha(1-\beta^{N})]=0\,,
\end{equation}
while the $N$th row of the eigenvalue equation is
\begin{equation}
b[A\alpha(\alpha^{N}-1)+B\beta(\beta^{N}-1)]=0\,.
\end{equation}
The two equation are exactly same as Eqs.~(\ref{c1}) and (\ref{cN}).

Now, in addition, the ($N+1$)st row of the eigenvalue equation reads
\begin{equation}
-d^*(v_1+v_2+\cdots+v_N)+(c-\lambda)v_{N+1}=0\,,
\end{equation}
and, by using the general solution, this can be reduced to
\begin{equation}
-d^*\left(A\frac{1-\alpha^N}{1-\alpha}+B\frac{1-\beta^N}{1-\beta}
-\frac{Nd\,v_{N+1}}{b+b^*-(a-\lambda)}\right)
+(c-\lambda)v_{N+1}=0\,.
\end{equation}

As in the previous section, we require that a nontrivial set
of $(A, B, v_{N+1})$ exists.
This leads to the following equation:
\begin{eqnarray}
& &\left|\begin{array}{ccc}
\beta(1-\alpha^{N})&\alpha(1-\beta^{N})&0\\
\alpha(\alpha^{N}-1)&\beta(\beta^{N}-1)&0\\
-d^*\frac{1-\alpha^N}{1-\alpha}&-d^*\frac{1-\beta^N}{1-\beta}&
\frac{N|d|^2}{b+b^*-(a-\lambda)}
+c-\lambda
\end{array}
\right|\nonumber \\
&=&(\beta^2-\alpha^2)(1-\alpha^{N})(\beta^{N}-1)
\left[\frac{N|d|^2}{b+b^*-(a-\lambda)}
+c-\lambda\right]=0
\,.
\end{eqnarray}
The second left-hand side of the equation should be proportional to
$D(\lambda)$, as for discussion in the previous section.
Because we have already known the normalization of
$b^N(1-\alpha^{N})(\beta^{N}-1)$, we conclude that
the characteristic polynomial in the present case is written by
\begin{equation}
D(\lambda)=b^N(1-\alpha^{N})(\beta^{N}-1)
\left[\frac{N|d|^2}{b+b^*-(a-\lambda)}
+c-\lambda\right]\,.
\end{equation}
Thus, the determinant of $H$ in this section is given by
\begin{eqnarray}
& &D(0)=\left(c-\frac{N|d|^2}{a-b-b^*}\right)
\left[\left(\frac{a+\sqrt{a^2-4|b|^2}}{2}\right)^{N}
+\left(\frac{a-\sqrt{a^2-4|b|^2}}{2}\right)^{N}-b^N-{b^*}^N\right]
\nonumber \\
& &\qquad\mbox{(for an extended periodic tridiagonal matrix)}\,.
\end{eqnarray}

\subsection{examples}

\subsubsection*{\textbullet~ $a=2r+s+l^2$,
$c=Ns+l^2$, $d=s$ and $b=r$ \cite{BLS,BHS}.%
\footnote{In this case,
\begin{equation}
\lambda_0=l^2\,,\quad\lambda_p=s+r\sin^2\frac{\pi p}{N}+l^2\quad
(p=1, 2, \ldots, N-1)\,,\quad
\lambda_N=(N+1)s+l^2\,.
\end{equation}
}
}

Suppose the $(N+1)\times (N+1)$ matrix \cite{BLS,BHS},
\begin{equation}
\Delta(r,s)=r\left(
\begin{array}{ccccccc}
2&-1&0&\cdots& &-1&0\\
-1&2&-1&\cdots& &0&0\\
0&-1&2&\cdots& &0&0\\
\vdots&\vdots&\vdots&\ddots& &\vdots&\vdots\\
 & & & &2&-1&0\\
-1&0&0&\cdots&-1&2&0 \\
0&0&0&\cdots&0&0&0
\end{array}
\right)+
s\left(
\begin{array}{ccccccc}
1&0&0&\cdots& &0&-1\\
0&1&0&\cdots& &0&-1\\
0&0&1&\cdots& &0&-1\\
\vdots&\vdots&\vdots&\ddots& &\vdots&\vdots\\
 & & & & & & \\
0&0&0&\cdots& &1&-1 \\
-1&-1&-1&\cdots& &-1&N
\end{array}
\right)\,,
\end{equation}
where $r$ and $s$ are constants.
The determinant of $H=\Delta(r,s)+l^2I_{N+1}$ is
\begin{equation}
\det H=D(0)=l^2\left(1+\frac{Ns}{s+l^2
}\right)
\left(\eta_+^{N}
+\eta_-^{N}-2r^N\right)\,,
\label{gw1}
\end{equation}
where
\begin{equation}
\eta_\pm\equiv\frac{1}{2}\left[2r+s+l^2\pm\sqrt{
\left(2r+s+l^2\right)^2-4r^2}\right]\,.
\label{gw2}
\end{equation}

\subsubsection*{\textbullet~ $a=3+l^2$,
$c=N+l^2$ and $b=d=1$. }

\begin{figure}[ht]
\centering
\includegraphics[height=5cm]
{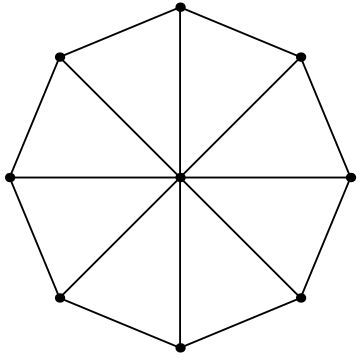}
\caption{%
$W_{9}$ as an example of a wheel graph.
}
\label{wheel}
\end{figure}

This is the previous case with $r=s=1$.
In this case, $H=\Delta(W_{N+1})+l^2I_{N+1}$,
where
\begin{equation}
\Delta(W_{N+1})\equiv\left(
\begin{array}{ccccccc}
3&-1&0&\cdots& &-1&-1\\
-1&3&-1&\cdots& &0&-1\\
0&-1&3&\cdots& &0&-1\\
\vdots&\vdots&\vdots&\ddots& &\vdots&\vdots\\
 & & & &3&-1&-1\\
-1&0&0&\cdots&-1&3&-1 \\
-1&-1&-1&\cdots&-1&-1&N
\end{array}
\right)\,
\end{equation}
is the graph Laplacian of the wheel graph (see FIG.~\ref{wheel}) with
$N+1$ vertices.
We do not repeat writing the expression of $H$,
which is given by Eq.~(\ref{gw1}) with Eq.~(\ref{gw2})
when $r=s=1$.

\subsubsection*{\textbullet~  $b=0$. }

\begin{figure}[ht]
\centering
\includegraphics[height=5cm]
{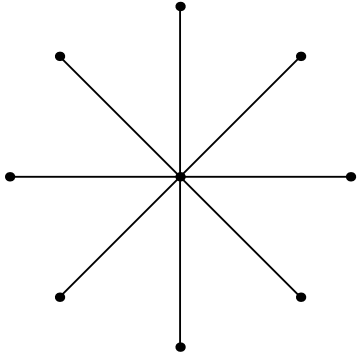}
\caption{%
$K_{1,8}$ as an example of a star graph.
}
\label{star}
\end{figure}

In this case, the determinant simply becomes as
\begin{equation}
\det H=D(0)=a^{N}c-{Na^{N-1}|d|^2}\,.
\end{equation}
Especially, $\Delta(0,1)$ is in this category, and can be written as
\begin{equation}
\Delta(0,1)=\left(
\begin{array}{ccccccc}
1&0&0&\cdots& &0&-1\\
0&1&0&\cdots& &0&-1\\
0&0&1&\cdots& &0&-1\\
\vdots&\vdots&\vdots&\ddots& &\vdots&\vdots\\
 & & & & & & \\
0&0&0&\cdots& &1&-1 \\
-1&-1&-1&\cdots& &-1&N
\end{array}
\right)=\Delta(K_{1,N})\,.
\end{equation}
This is the graph Laplacian of a star graph $K_{1,N}$ (FIG.~\ref{star}).
The eigenvalues of $\Delta(K_{1,N})$ are known as
\begin{equation}
\lambda_0=0\,,\quad\lambda_p=1\quad
(p=1, 2, \ldots, N-1)\,,\quad\lambda_N=N+1.
\end{equation}
The determinant of $H=\Delta(K_{1,N})+l^2I_{N+1}$ is
\begin{equation}
\det H=D(0)=l^2\left(1+\frac{N}{1+l^2
}\right)
\left(1+l^2\right)^N\,.
\end{equation}

\section{Free energy on a graph} 
\label{sec6}

In this section, we consider applications of the results on determinants
for studying discrete systems.

We first consider $N$ scalar degrees of freedom 
and define the action as follows:
\begin{equation}
S=\frac{1}{2}\left(\frac{N^2}{L^2}\sum_{k=1}^N\sum_{k'=1}^N
\phi_k\Delta(C_N)_{kk'}\phi_{k'}+
\mu^2\sum_{k=1}^N\phi_k^2\right)\,,
\label{SC}
\end{equation}
where $\Delta(C_N)$ is the graph Laplacian for $C_N$ and
\begin{equation}
\mu\equiv\frac{2N}{L}\sinh\frac{mL}{2N}\,,
\label{choice}
\end{equation}
where $L$ and $m$ are constants.

Then, the Gaussian free energy on $C_N$ \cite{KSJPA} is obtained using
Eq.~(\ref{dcyc}) as
\begin{equation}
F_{C_N}=\frac{1}{2}\ln\left[\det
\left(\Delta(C_N)+4\sinh^2\frac{mL}{2N}I_N\right)\right]=\ln\left(
2\sinh\frac{mL}{2}\right)+const.\,.
\end{equation}
This is interesting because the action (\ref{SC}) can be rewritten as
\begin{equation}
S=\frac{1}{2}\left[\frac{N^2}{L^2}\sum_{k=1}^N
(\phi_k-\phi_{k+1})^2+
\mu^2\sum_{k=1}^N\phi_k^2\right]\,,
\end{equation}
under the `periodic' condition, $\phi_{N+1}\equiv\phi_1$.
A continuum limit, $a_0\equiv L/N\rightarrow 0$, enforces
$\frac{\phi_{k+1}-\phi_k}{a_0}\rightarrow \partial_x\phi$, where $x$ is a
coordinate of one dimension with periodicity $x+L\sim x$.
Therefore, we can find that the one-loop free energy of a real scalar
field
$\phi(x)$ with mass
$m$ on a circle ($S_1$) with circumference $L$ governed by the action
\begin{equation}
S=\frac{1}{2}\int dx \left[(\partial_x\phi)^2+m^2\phi^2\right]\,
\label{SS}
\end{equation}
takes the form
\begin{equation}
F_{S_1}=\ln\left(
2\sinh\frac{mL}{2}\right)\,,
\end{equation}
after some regularization \cite{KSJPA,Monin}.
Note that since
\begin{equation}
2\sinh\frac{mL}{2}=mL\frac{\prod_{n=1}^\infty
\left(\frac{4\pi^2n^2}{L^2}+m^2\right)}{\prod_{n=1}^\infty
\left(\frac{4\pi^2n^2}{L^2}\right)}\,,
\end{equation}
we find that the eigenvalues of $-\partial_x^2+m^2$, where
$-\partial_x^2$ is the one-dimensional Laplacian on
$S_1$, are shown by
\begin{equation}
\frac{4\pi^2n^2}{L^2}+m^2\quad (n \mbox{ is an integer})\,.
\end{equation}

\bigskip

Similarly, we can consider the other matrices. For example,
the action for complex scalar fields defined as
\begin{eqnarray}
S&=&\frac{1}{2}\left(\frac{N^2}{L^2}\sum_{k=1}^N\sum_{k'=1}^N
\phi^\dagger_k{\Delta(C_{N},\chi)}_{kk'}\phi_{k'}+
\mu^2\sum_{k=1}^N|\phi_k|^2\right)\nonumber \\
&=&\frac{1}{2}\left[\frac{N^2}{L^2}\sum_{k=1}^N
|\phi_k-e^{i\chi}\phi_{k+1}|^2+
\mu^2\sum_{k=1}^N|\phi_k|^2\right]\,,
\end{eqnarray}
leads to the free energy
\begin{eqnarray}
F_{C_N,\chi}&=&\ln\left[\det
\left(\Delta(C_N,\chi)+4\sinh^2\frac{mL}{2N}I_N\right)\right]\nonumber \\
&=&
\ln\left(
4\sinh^2\frac{Nz}{2}+4\sin^2\frac{N\chi}{2}\right)+const.\,.
\end{eqnarray}
Here we will avoid repeated discussion, and only note that 
the eigenvalue spectrum of the continuum limit of this case
is given by 
\begin{equation}
\frac{(2\pi n-\chi)^2}{L^2}+m^2\quad (n \mbox{ is an integer})\,.
\end{equation}

\bigskip

Continuum limits exist also in other some cases.

\bigskip

The large $N$ limit of the determinant of $\Delta_{DD}+\mu^2I_N$ 
(according to Eq.~(\ref{dDD}))
becomes 
\begin{equation}
\frac{\sinh mL}{\sinh\frac{mL}{N}}\rightarrow N\frac{\sinh
mL}{{mL}}\,,
\end{equation}
which coincides with the result of the example stated in Sec.~\ref{sec2}
up to the constant. We find that the continuum limit corresponds to the
system of massive scalar field in a line $0\le x\le L$ with
Dirichlet-Dirichlet boundary conditions at its ends.

\bigskip

The large $N$ limit of the determinant of $\Delta_{DN}+\mu^2I_N$ 
(according to Eq.~(\ref{dDN}))
becomes simply
\begin{equation}
\cosh mL+2\sinh \frac{mL}{N}\sinh mL\rightarrow \cosh mL\,.
\end{equation}
A comparison to a known mathematical relation
\begin{equation}
\cosh
mL=\frac{\prod_{n=0}^\infty\left[\frac{\pi^2}{L^2}\left(n+\frac{1}{2}\right)^2
+m^2\right]}{\prod_{n=0}^\infty\left[\frac{\pi^2}{L^2}\left(n+\frac{1}{2}
\right)^2
\right]}
\end{equation}
leads to the conclusion that the continuum limit of spectrum is given by
$\frac{\pi^2}{L^2}\left(n+\frac{1}{2}\right)^2+m^2$, thus
the boundary conditions of the system is Dirichlet-Neumann condition.

\bigskip

Finally, the determinant of $\Delta(P_N)+\mu^2I_N$ 
(according to Eq.~(\ref{dNN}))
is
\begin{equation}
2\tanh\frac{mL}{2N}\sinh mL\,.
\end{equation}
Since the free energy is proportional to the logarhithm of this,
we drop the $N$ dependent term (which is log divergent if
$N\rightarrow\infty$). The boundary conditions of the continuum system is
Neumann-Neumann condition (which can be judged from the existence of a
zero mode).

\bigskip

In the next section, we will consider the way to obtain
one-loop vacuum energy of scalar field theory with mass matrix
required by structure of a theory space 
with four dimensional spacetime.

\section{Vacuum energy from a theory space}
\label{sec7}

\subsection{formulation}

One-loop vacuum energy density in quantum field theory can be derived from
the functional determinants \cite{Dunne}. In the present paper, we only
consider scalar field theories for simplicity.
As seen in the previous section, $N$-scalar field theory can resemble
compactification of a dimension. This is the key idea of the dimensional
deconstruction \cite{ACG,HPW,HL}. The structure of the theory space
is determined by the quadratic term of fields, i.e., the mass matrix.
Suppose that a mass matrix (precisely, the $(mass)^2$ matrix)
$\Delta/a_0^2$ is given (in other words, a theory space is given). The
eigenvalues of $\Delta$ are denoted by $\lambda_p$, as previously.
Then, using the characteristic polynomial $D(\lambda)=\prod_p
(\lambda_p-\lambda)$, one-loop vacuum energy density for real scalar
fields is calculated by
\begin{equation}
V=\frac{1}{2}\int\frac{d^4l}{(2\pi)^4}\ln
\frac{\det[\Delta/a_0^2+l^2]}{\det[\Delta/a_0^2+l^2+M^2]}=\frac{1}{2a_0^4}(2\pi^2)\int_0^\infty\frac{l^3dl}{(2\pi)^4}\ln
\frac{D(-l^2)}{D(-l^2-M^2a_0^2)}\,,
\label{formalve}
\end{equation}
where we used $M$ of the Pauli-Villars regularization,
which is considered to be $M\rightarrow\infty$.
The constant $a_0$ illustrates an overall scale in the theory space,
i.e., related to  mass scale of new physics via $a_0\sim
m_{new~physics}^{-1}$.

In practice, regularization is an art of assembly of mathematical
techniques. We adopt here the following approach.
A physical value of the vacuum energy should be determined independently
of the unphysical $M$ and the UV divergence must be subtracted in the
expression of it. Thus, we consider, in the denominator in log in
Eq.~(\ref{formalve}), as
\begin{equation}
D(-l^2-M^2a_0^2)\Rightarrow \mbox{Asymptotic form of } D(-l^2)
\mbox{ when } l^2\rightarrow\infty\,.
\end{equation}
Further, if the theory contains the $N$ (scalar) fields, the integrand
of the most divergent part should be proportional to $N$.
Thus, we extract the part of
$\propto (l^2)^N\subset D(-l^2)$ for large $l^2$.

\subsection{dimensional deconstruction of a circle}

A concrete example is in order.
We consider a theory space associated with $\Delta(C_N,\chi)$.
This model has widely been studied by many authors \cite{ACG,HPW,HL,KSS}. 
We have already obtained $\det
\left[\Delta(C_N,\chi)+4\sinh^2\frac{z}{2}\,I_N\right]$ ($\propto
D(-l^2)$, in the present case) in Eq.~(\ref{dcycchi}). The asymptotic
behavior can be found as
\begin{eqnarray}
& &\det\left[
\Delta(C_N,\chi)+4\sinh^2\frac{z}{2}\,I_N\right]\nonumber \\
&=&
\left|e^{N(z+i\chi)/2}-e^{-N(z+i\chi)/2}\right|^2\rightarrow
\left|e^{N(z+i\chi)/2}\right|^2 (=e^{Nz})\quad (z\rightarrow \infty).
\end{eqnarray}
Thus, in our regularization scheme,%
\footnote{We now consider complex scalar fields.}
\begin{equation}
V(\chi)=\frac{1}{a_0^4}(2\pi^2)\int_0^\infty\frac{dz}{(2\pi)^4}
8\sinh^3\frac{z}{2}\cosh\frac{z}{2}\ln
\frac{\left|e^{N(z+i\chi)/2}-
e^{-N(z+i\chi)/2}\right|^2}{\left|e^{N(z+i\chi)/2}\right|^2}\,,
\end{equation}
where we set $l=2\sinh\frac{z}{2}$.
Now, the integration can be done by elementary methods as
\begin{eqnarray}
V(\chi)&=&\frac{2\pi^2}{a_0^4}\int_0^\infty\frac{dz}{(2\pi)^4}
4\sinh^2\frac{z}{2}\sinh z\left[\ln\left(
1-e^{-N(z+i\chi)}\right)+\ln\left(
1-e^{-N(z-i\chi)}\right)
\right]\nonumber \\
&=&-\frac{1}{16\pi^2a_0^4}\sum_{n=1}^\infty\int_0^\infty dz
(e^\frac{z}{2}-e^{-\frac{z}{2}})^2(e^z-e^{-z}
)\left[\frac{e^{-Nn(z+i\chi)}}{n}+\frac{e^{-Nn(z-i\chi)}}{n}
\right]\nonumber \\
&=&-\frac{1}{8\pi^2a_0^4}\sum_{n=1}^\infty\frac{\cos
Nn\chi}{n}\int_0^\infty dz (e^{2z}-2e^z+2e^{-z}-e^{-2z})
e^{-Nnz}\nonumber \\
&=&-\frac{1}{8\pi^2a_0^4}\sum_{n=1}^\infty\frac{\cos
Nn\chi}{n}\left(\frac{1}{Nn-2}-\frac{2}{Nn-1}+\frac{2}{Nn+1}-
\frac{1}{Nn+2}\right)\nonumber \\
&=&-\frac{1}{8\pi^2a_0^4}\sum_{n=1}^\infty\frac{\cos
Nn\chi}{n}\left(\frac{4}{N^2n^2-4}-\frac{4}{N^2n^2-1}\right)\nonumber \\
&=&-\frac{3}{2\pi^2a_0^4}\sum_{n=1}^\infty\frac{\cos
Nn\chi}{n(N^2n^2-1)(N^2n^2-4)}\,.
\end{eqnarray}
This result exactly coincides with the known result
\cite{ACG,HPW,HL,KSS}.%
\footnote{It is notable that the regularized vacuum energy is
 $O(N^{-4})$, while the subtracted part whose integrand
including $\ln e^{Nz}=Nz$ is proportional to $N$.}
Incidentally, for large $N$, 
\begin{equation}
V(\chi)\sim-\frac{3}{2\pi^2(Na_0)^4}\sum_{n=1}^\infty\frac{\cos
Nn\chi}{n^5}\,,\quad
V(0)\sim-\frac{3\zeta(5)}{2\pi^2(Na_0)^4}\,,
\end{equation}
where $\zeta(z)$ is the Riemann's zeta function.
We find that there exists a ``continuum limit'', $N\rightarrow\infty$
as $Na_0$ and $N\chi$ are fixed.

\subsection{the clockwork theory}

Next, we turn to consider the theory space of the clockwork theory
\cite{CI,KR,GM,FPRT} for real scalar fields.
The action is
\begin{equation}
S=\frac{1}{2}\int d^4x
\left[\sum_{k=1}^N(\partial_\mu\phi_k)^2+m^2\sum_{k=1}^{N-1}
(\phi_k-q\phi_{k+1})^2\right]\,,
\end{equation}
where $m=a_0^{-1}$.
Thus, the relevant matrix determinant is given as Eq.~(\ref{cw}).
The subtraction of UV divergence is subtle because of the complicated
form of the determinant in this case. 
We separate the vacuum energy density into three parts, such as
$V=V_N(q)+V_0+NV_1$.
Here, $V_N(q)$ is a finite part ,
\begin{equation}
V_N(q)=\frac{1}{2a_0^4}(2\pi^2)\int_0^\infty\frac{l^3dl}{(2\pi)^4}
\ln
\left[1-\left(\frac{\gamma_-}{\gamma_+}\right)^N\right]\,,
\end{equation}
where $\gamma_\pm$ is given by Eq.~(\ref{gam}).
This will be of order of $O(N^{-4})$ as in the previous case and thus
will have a continuum limit in vacuum energy density.

The change of the integration variable $\cosh y=\frac{\l^2+1+q^2}{2q}$
makes the integration simple. Then, we can rewrite $V_N(q)$ as
\begin{equation}
V_N(q)=\frac{1}{2a_0^4}(2\pi^2)\int_{|\ln q|}^\infty\frac{dy}{(2\pi)^4}
q\sinh y (2q\cosh y-1-q^2)\ln
\left(1-e^{-2Ny}\right)\,,
\end{equation}
and we get the form with infinite summations,
\begin{eqnarray}
V_N(q)&=&-\frac{q^2}{16\pi^2a_0^4}\sum_{n=1}^\infty
\frac{q^{-2Nn}}{2n}
\left[\frac{q^2}{(2Nn-2)(2Nn-1)}
\right.\nonumber \\
&
&\qquad\left.-2\frac{1}{(2Nn-1)(2Nn+1)}+\frac{q^{-2}}{(2Nn+1)(2Nn+2)}\right]
\quad (q\ge 1)\,,
\end{eqnarray}
\begin{eqnarray}
V_N(q)&=&-\frac{q^2}{16\pi^2a_0^4}\sum_{n=1}^\infty
\frac{q^{2Nn}}{2n}
\left[\frac{q^{-2}}{(2Nn-2)(2Nn-1)}
\right.\nonumber \\
&
&\qquad\left.-2\frac{1}{(2Nn-1)(2Nn+1)}+\frac{q^2}{(2Nn+1)(2Nn+2)}\right]
\quad (q< 1)\,.
\end{eqnarray}
Note that $q^{-2}V_N(q)=q^2V_N(q^{-1})$.
\begin{figure}[ht]
\centering
\includegraphics
{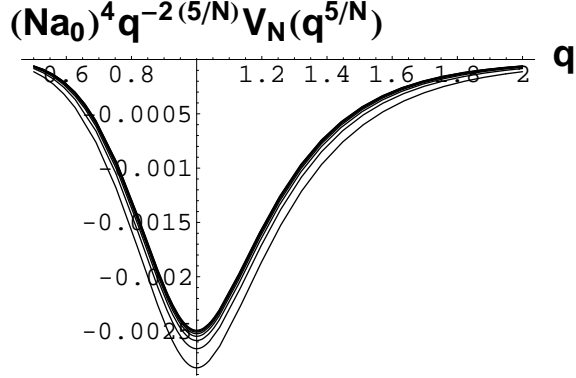}
\caption{%
The numerical results for $(Na_0)^4q^{-2(5/N)}V_N(q^{5/N})$ 
as functions of $q$ for $N=3, 4,\ldots, 10$, from lower to upper curves.
}
\label{vq}
\end{figure}
The numerical results for $(Na_0)^4q^{-2(5/N)}V_N(q^{5/N})$ are shown in
FIG.~\ref{vq} for $N=3, 4,\ldots, 10$.
These curves indicate that there is a continuum limit
$N\rightarrow\infty$, while $Na_0$ and $q^{1/N}$ are fixed constants.
If we can treat $q$ as a dynamical variable,
the effective potential of $q$ seems to have a minimum at $q\sim 1$
for large $N$, where the mass matrix simply becomes the graph Laplacian of
$P_N$. Note also that $V_N(q)\rightarrow 0$ both for $q\rightarrow 0$ and
for $q\rightarrow\infty$.

We now estimate the separated contributions.
They are written as
\begin{equation}
V_{0}=\frac{1}{2a_0^4}(2\pi^2)\int_0^\infty\frac{l^3dl}{(2\pi)^4}
\ln
\frac{l^2}{\sqrt{(1-q^2)^2+2(1+q^2)l^2+l^4}}\,,
\end{equation}
and
\begin{eqnarray}
V_{1}&=&\frac{1}{2a_0^4}(2\pi^2)\int_0^\infty\frac{l^3dl}{(2\pi)^4}
\,\ln\gamma_+\nonumber \\
&=&\frac{1}{2a_0^4}(2\pi^2)\int_0^\infty\frac{l^3dl}{(2\pi)^4}
\ln
\left[\frac{1+q^2+l^2+\sqrt{(1-q^2)^2+2(1+q^2)l^2+l^4}}{2}\right]\,.
\end{eqnarray}

As for $V_0$, if we use the standard formula of derivation of the
Coleman-Weinberg potential
\begin{equation}
\left.\frac{1}{2}\int\frac{d^4l}{(2\pi)^4}\ln(l^2+M^2)\right|_{regularized}
=\frac{M^4}{64\pi^2}\ln
M^2\,,
\end{equation}
to regularize $V_0$, aside from the contribution of a zero mode (as
$\ln l^2$ in the integrand), we find
\begin{equation}
V_{0}=-\frac{1}{2(64\pi^2)a_0^4}
[(1-q)^4\ln (1-q)^2+(1+q)^4\ln (1+q)^2]\,.
\end{equation}
It is notable that this contribution is equivalent to subtraction of the
half of vacuum energy densities due to scalar fields with mass squared
$(1-q)^2/a_0^2$ and $(1+q)^2/a_0^2$. The UV divergence of this part can
be regarded to be canceled by the zero-mode contribution.

On the other hand, for the complicated form of a genuine divergent
contribution of $NV_1$, we introduce a cut-off
$\Lambda$ in the integration over $l$ and find
\begin{eqnarray}
NV_{1}&=&\frac{N}{64\pi^2a_0^4}\left[
\Lambda^4\left(\ln\Lambda^2-\frac{1}{2}\right)+2(1+q^2)\Lambda^2-(1+4q^2+q^4)\ln
\frac{2\Lambda^2}{1+q^2+|1-q^2|}\right.
\nonumber
\\ & &\left.+(1+q^2)^2-\frac{3}{2}(1+q^2)|1-q^2|\right]\,.
\end{eqnarray}
The quartic divergence seems to be independent of the structure of
the mass matrix and the quadratic divergence is proportional to
the trace of the mass matrix.

\subsection{latticization of a disk}

The matrix $\Delta(r,s)$ is used in \cite{BLS,BHS}
as a latticization of a disk.
Using the result of Eqs.~(\ref{gw1}) and (\ref{gw2}), one-loop vacuum
energy density of scalar field theory with mass matrix
$\Delta(r,s)/a_0^2$ can be written formally as
\begin{eqnarray}
& &V=\frac{2\pi^2}{2a_0^4}\int_0^\infty\frac{l^3dl}{(2\pi)^4}
\ln\left[ l^2\left(1+\frac{Ns}{s+l^2
}\right)
\left(\eta_+^{N}
+\eta_-^{N}-2r^N\right)\right]\nonumber \\
& &=V_N(r,s)+V_0+NV_1\,,
\end{eqnarray}
where
\begin{eqnarray}
V_N&=&\frac{2\pi^2}{2a_0^4}\int_0^\infty\frac{l^3dl}{(2\pi)^4}
\ln
\left[1
+\left(\frac{\eta_-}{\eta_+}\right)^{N}-2\left(\frac{r}{\eta_+}
\right)^N\right]\nonumber \\
&=&\frac{2\pi^2}{2a_0^4}\int_0^\infty\frac{l^3dl}{(2\pi)^4}
\, 2\ln
\left[1-\left(\frac{r}{\eta_+}
\right)^N\right]\,,
\end{eqnarray}
with
\begin{equation}
\eta_+\equiv\frac{1}{2}\left[2r+s+l^2+\sqrt{
\left(2r+s+l^2\right)^2-4r^2}\right]\,,
\end{equation}
\begin{equation}
V_0=\frac{2\pi^2}{2a_0^4}\int_0^\infty\frac{l^3dl}{(2\pi)^4}
\ln\left[l^2\left(1+\frac{Ns}{s+l^2}\right)\right]
\,,
\end{equation}
and
\begin{eqnarray}
V_1&=&\frac{2\pi^2}{2a_0^4}\int_0^\infty\frac{l^3dl}{(2\pi)^4}
\ln\eta_+\nonumber \\
&=&\frac{2\pi^2}{2a_0^4}\int_0^\infty\frac{l^3dl}{(2\pi)^4}
\ln\frac{2r+s+l^2+\sqrt{
\left(2r+s+l^2\right)^2-4r^2}}{2}
\,.
\end{eqnarray}

A finite part $V_N$ can be rewritten as
\begin{equation}
V_N(s/r)
=\frac{4\pi^2 r^2}{2a_0^4}\int_0^\infty\frac{l^3dl}{(2\pi)^4}
\, \ln
\left[1-\left(\frac{2}{2+s/r+l^2+\sqrt{
\left(2+s/r+l^2\right)^2-4}}
\right)^N\right]\,.
\end{equation}
Furthermore, introducing new variables
$s/r=2\sin^2\frac{u}{2}$ and $\cosh y=l^2/2+\cosh u$,
we find
\begin{eqnarray}
V_N(u)
&=&\frac{4\pi^2 r^2}{2a_0^4}\int_u^\infty\frac{dy}{(2\pi)^4}
\, 2\sinh y(\cosh y-\cosh u)\ln
\left[1-e^{-Ny}\right]\nonumber \\
&=&-\frac{r^2}{8\pi^2a_0^4}\sum_{n=1}^\infty
\frac{e^{-Nnu}}{n}
\left[\frac{e^{2u}}{(Nn-2)(Nn-1)}
\right.\nonumber \\
&
&\qquad\left.-2\frac{1}{(Nn-1)(Nn+1)}+\frac{e^{-2u}}{(Nn+1)(Nn+2)}\right]
\,.
\end{eqnarray}

\begin{figure}[ht]
\centering
\includegraphics
{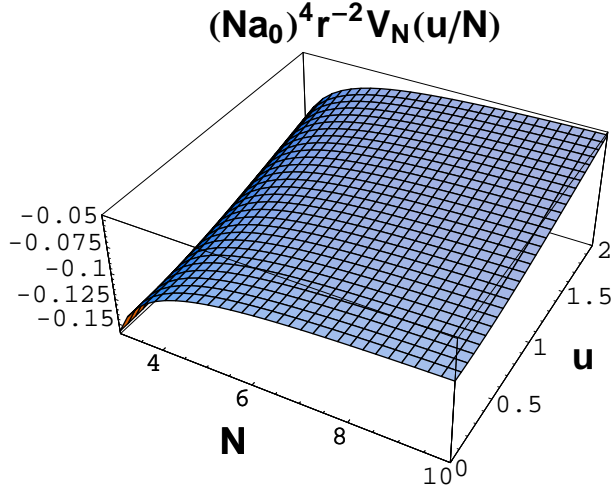}
\caption{%
The numerical results for $(Na_0)^4r^{-2}V_N(u/N)$ 
as a function of $N$ and $u$.
}
\label{wheelve}
\end{figure}

The numerical result of $(Na_0)^4r^{-2}V_N(u/N)$ is plotted as a
function of $N$ and $u$ in FIG.~\ref{wheelve}, where we treat $N$ as a
continuous parameter.
We find that
$\lim_{N\rightarrow\infty}(Na_0)^4r^{-2}V_N(0)=-\frac{3\zeta(5)}{4\pi^2}$,
while we find no other limiting case for general $r$ and $s$, i.e.,
no precise continuum limit exists in general cases.

We now turn to consider the other part of the vacuum energy.
For $V_0$, using similar estimation as in the previous subsection,
we obtain, up to the zero-mode contribution,
\begin{equation}
V_0=\frac{1}{64\pi^2a_0^4}s^2\Bigl\{(N+1)^2\ln[(N+1)^2s^2]-\ln s^2\Bigr\}
\,,
\end{equation}
which is equivalent to the contribution of a scalar field with mass
squared $(N+1)^2s^2/a_0^2$ minus the contribution of a scalar field with mass
squared $s^2/a_0^2$. The UV divergence is canceled in this two
contributions.

The divergent part is analyzed by using the cut-off $\Lambda$ and is
found to be
\begin{eqnarray}
NV_{1}&=&\frac{N}{64\pi^2a_0^4}\left[
\Lambda^4\left(\ln\Lambda^2-\frac{1}{2}\right)+2(2r+s)
\Lambda^2-(6r^2+4rs+s^2)\ln
\frac{2\Lambda^2}{2r+s+\sqrt{s(4r+s)}}\right.
\nonumber
\\ & &\left.+(2r+s)^2-\frac{3}{2}(2r+s)\sqrt{s(4r+s)}\right]\,.
\end{eqnarray}
Again, we find that the quartic divergence is independent of the mass
matrix and the quadratic divergence is proportional to the trace of
the mass matrix.%
\footnote{Note that, remembering the one zero-mode contribution
 separated from $V_0$, the quartic divergence is found to be proportional
to
$N+1$.}

In the next section, we will exhibit one more example of calculation
of one-loop vacuum energy density for a slightly complicated theory space.

\section{Some other examples of vacuum energy}
\label{sec8}

Using the additional formulas on determinants,
we can further obtain determinants of various matrices.
In this section, we show some other examples below.

\subsection{adding an edge with a vertex to each vertex of a graph}

Let $\Delta_N$ be an $N\times N$ Hermitian matrix and define a
$2N\times 2N$ matrix $\Delta_{2N}$ as follows:
\begin{equation}
\Delta_{2N}\equiv
\left(
\begin{array}{cc}
\Delta_N+I_N&-I_N\\
-I_N&I_N
\end{array}
\right)\,,
\end{equation}
where $I_N$ is an $N$ dimensional identity marix.
In particular, if $\Delta_N$ is the graph Laplacian of a graph $G$,
$\Delta_{2N}$ is the graph Laplacian of the graph generated by adding
an edge with a vertex to every vertices of $G$.

Then, the formula on deteminants 
\begin{equation}
\det\left(
\begin{array}{cc}
A&B\\
C&D
\end{array}
\right)=\det(A-BD^{-1}C)\det (D)
\end{equation}
tells us that
\begin{equation}
\det(\Delta_{2N}-\lambda I_{2N})=(1-\lambda)^N
\det\left(\Delta_N-\frac{1-(1-\lambda)^2}{1-\lambda}I_N\right)\,.
\end{equation}

Therefore, if $D_N(\lambda)\equiv
\prod_{p=1}^N(\lambda_p-\lambda)=\det(\Delta_N-\lambda I_N)$, where
$\{\lambda_p\}$ are eigenvalues of $\Delta_N$, is known,
$D_{2N}(\lambda)\equiv\det(\Delta_{2N}-\lambda I_{2N})$
is obtained as
\begin{equation}
D_{2N}(\lambda)=(1-\lambda)^ND_N\left(\frac{1-(1-\lambda)^2}{1-\lambda}\right)\,.
\end{equation}

\begin{figure}[ht]
\centering
\includegraphics[height=5cm]
{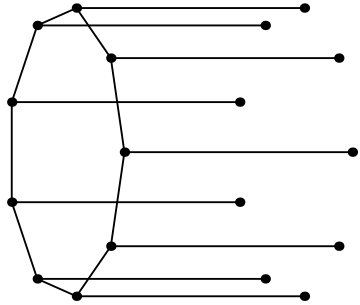}
\caption{%
The graph generated by adding
an edge with a vertex to every vertices of $C_9$.}
\label{sung}
\end{figure}

For example, we will calculate vacuum energy density of the scalar field
theory with mass matrix $\Delta_{2N}/a_0^2$, where $\Delta_{2N}$ is
generated from
$\Delta_N=\Delta(C_N)$, i.e., $\Delta_{2N}$ is the graph Laplacian of the
graph shown in FIG.~\ref{sung}.

In this case, after some manipulation, we get
\begin{equation}
D_{2N}(\lambda)=\alpha(\lambda)^N\left[1-
\left(\frac{1-\lambda}{\alpha(\lambda)}\right)^N\right]^2\,,
\end{equation}
where
\begin{equation}
\alpha(\lambda)\equiv\frac{\lambda^2-4\lambda+2+
\sqrt{\lambda(\lambda^3-8\lambda^2+16\lambda-8)}}{2}\,.
\end{equation}

Now, we can obtain the vacuum energy density in this theory
by utilizing $\ln D_{2N}(-l^2)$, as in the previous section. 
We separate the finite and divergent parts of vacuum energy density
as
\begin{equation}
V_N=\frac{1}{2a_0^4}(2\pi^2)\int_0^\infty\frac{l^3dl}{(2\pi)^4}2\ln
\left[1-
\left(\frac{1+l^2}{\alpha(-l^2)}\right)^N\right]\,,
\end{equation}
and
\begin{equation}
V_1=\frac{1}{2a_0^4}(2\pi^2)\int_0^\infty\frac{l^3dl}{(2\pi)^4}
\ln\alpha(-l^2)\,.
\end{equation}

\begin{figure}[ht]
\centering
\includegraphics
{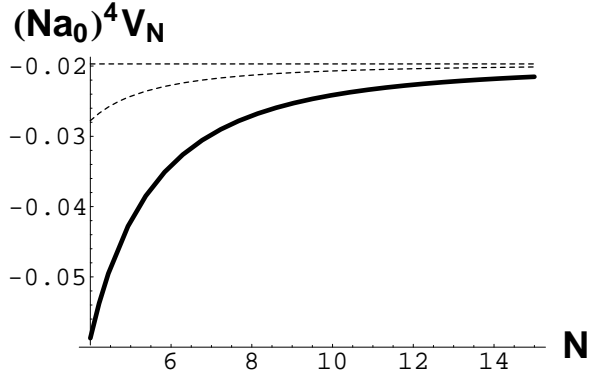}
\caption{%
The numerical value of $(Na_0)^4V_N$ for the model in this
subsection as a function of $N$.
The dotted lines indicate $\frac{1}{4}(Na_0)^4V_N$, where $V_N$ is the
vacuum energy density in the real scalar theory whose mass matrix is
$\Delta(C_N)/a_0^2$, and the constant $-\frac{3\zeta(5)}{16\pi^2}$.}
\label{sungve}
\end{figure}

The numerical result of $(Na_0)^4V_N$ in the present case is shown in
FIG.~\ref{sungve}, where $N$ is treated as a continuous parameter. In the
limit of
$N\rightarrow
\infty$,
$(Na_0)^4V_N$ approches
$-\frac{3\zeta(5)}{16\pi^2}$, which is quarter of the value of the large
$N$ limit of $(Na_0)^4V_N$ in the case of the real scalar theory based on
the graph Laplacian $\Delta(C_N)$. The divergent part $NV_1$ can be
estimated, because $\alpha(-l^2)\sim l^4$ for large $l$, as
\begin{equation}
NV_1=\frac{2N}{64\pi^2a_0^4}\Lambda^4\left(\ln\Lambda^2-\frac{1}{2}
\right)+\frac{N\Lambda^2}{8\pi^2}+\cdots\,.
\end{equation}
The leading term is proportional to the number of real scalar
fields, as expected. The quadratic divergence is proportional to the
trace of the mass matrix.

\subsection{the graph Cartesian products $G\times P_2$}

Let $\Delta_N$ be an $N\times N$ Hermitian matrix and define a
$2N\times 2N$ matrix $\hat\Delta_{2N}$ as follows:
\begin{equation}
\hat{\Delta}_{2N}\equiv
\left(
\begin{array}{cc}
\Delta_N+I_N&-I_N\\
-I_N&\Delta_N+I_N
\end{array}
\right)\,.
\end{equation}
In particular, if $\Delta_N$ is the graph Laplacian of a graph $G$,
$\hat{\Delta}_{2N}$ is the graph Laplacian of the graph Cartesian product
$G\times P_2$.%
\footnote{The graph Cartesian product of $G_1$ and $G_2$ is also often
written as
$G_1\Box G_2$.}

Then, the use of the formula on deteminants 
\begin{equation}
\det\left(
\begin{array}{cc}
A&B\\
C&D
\end{array}
\right)=\det(A-BD^{-1}C)\det (D)=\det(AD-BC)\,,
\end{equation}
provided that $[C, D]=0$,
leads to
\begin{eqnarray}
\det(\hat{\Delta}_{2N}-\lambda I_{2N})&=&
\det\left([\Delta_N+(1-\lambda)I_N]^2-I_N\right)\nonumber \\
&=&
\det\left(\Delta_N-\lambda I_N\right)\cdot
\det\left(\Delta_N+(2-\lambda)I_N\right)\,.
\end{eqnarray}
Therefore, if $D_N(\lambda)\equiv
\prod_{p=1}^N(\lambda_p-\lambda)=\det(\Delta_N-\lambda I_N)$, where
$\{\lambda_p\}$ are eigenvalues of $\Delta_N$, is known,
$\hat{D}_{2N}(\lambda)\equiv\det(\hat\Delta_{2N}-\lambda I_{2N})$
is obtained as
\begin{equation}
\hat{D}_{2N}(\lambda)=D_N(\lambda)D_N(\lambda-2)\,.
\end{equation}

\begin{figure}[ht]
\centering
\includegraphics[height=5cm]
{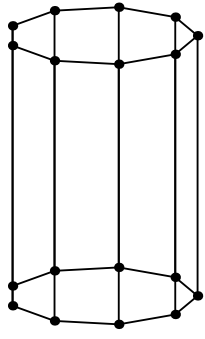}
\caption{%
The graph Cartesian product $C_9\times P_2$, or the prism graph $Y_9$.}
\label{prism}
\end{figure}

The vacuum energy density of the scalar field
theory with mass matrix $\hat\Delta_{2N}/a_0^2$, where $\hat\Delta_{2N}$
is generated from
$\Delta_N=\Delta(C_N)$, i.e., $\hat\Delta_{2N}$ is the graph Laplacian of
the graph Cartesian product $C_N\times P_2$, called as the prism graph
$Y_N$. We show the graph $Y_9$ in FIG.~\ref{prism}.

We can obtain the vacuum energy density in this theory
by utilizing $\ln \hat{D}_{2N}(-l^2)$. 
We separate the finite and divergent parts of vacuum energy density
as
\begin{eqnarray}
V_N&=&\frac{1}{16\pi^2a_0^4}\int_0^\infty{l^3dl}\cdot 2
\left\{\ln
\left[1-
\left(\frac{2}{2+l^2+\sqrt{(2+l^2)^2-4}}\right)^N\right]\right.\nonumber
\\ & &\qquad\qquad\qquad\qquad\left.+
\ln
\left[1-
\left(\frac{2}{4+l^2+\sqrt{(4+l^2)^2-4}}\right)^N\right]
\right\}\,,
\end{eqnarray}
and
\begin{equation}
V_1=\frac{1}{16\pi^2a_0^4}\int_0^\infty{l^3dl}
\left[\ln\frac{2+l^2+\sqrt{(2+l^2)^2-4}}{2}+\ln
\frac{4+l^2+\sqrt{(4+l^2)^2-4}}{2}\right]\,.
\end{equation}

\begin{figure}[ht]
\centering
\includegraphics
{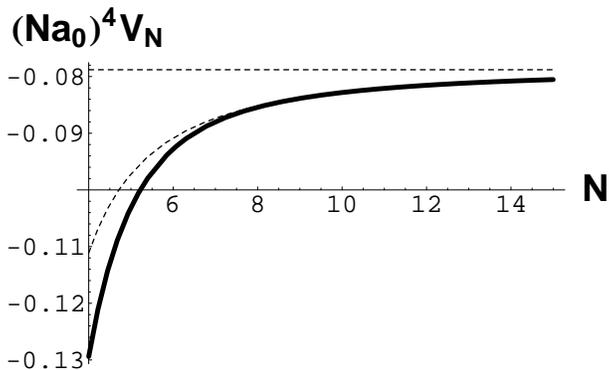}
\caption{%
The numerical value of $(Na_0)^4V_N$ for the model whose theory space is
associated with $Y_N$ as a function of $N$.
The dotted lines indicate $(Na_0)^4V_N$, where $V_N$ is the
vacuum energy density in the real scalar theory whose mass matrix is
$\Delta(C_N)/a_0^2$, and the constant $-\frac{3\zeta(5)}{4\pi^2}$.}
\label{prismve}
\end{figure}

The numerical result of $(Na_0)^4V_N$ in the present case is shown in
FIG.~\ref{prismve}, where $N$ is treated as a continuous parameter. In the
limit of
$N\rightarrow
\infty$,
$(Na_0)^4V_N$ approches the vacuum energy density of the model 
associated with $C_N$ and
$-\frac{3\zeta(5)}{4\pi^2}$. The divergent part $NV_1$ can be
estimated as
\begin{equation}
NV_1=\frac{2N}{64\pi^2a_0^4}\Lambda^4\left(\ln\Lambda^2-\frac{1}{2}
\right)+\frac{3N\Lambda^2}{16\pi^2}+\cdots\,.
\end{equation}
The leading term is proportional to the number of real scalar
fields, as expected.
The quadratic divergence is proportional to the
trace of the mass matrix.

\section{Conclusion}
\label{concl}

In the present paper, we showed the method of obtaining the determinant
of repetitive tridiagonal matrices with concrete examples.
The concept of the method is similar to the Gel'fand--Yaglom method of
obtaining functional determinants for differential operators.

The repetitive matrices as mass matrices are widely considered in
modern models in particle physics, in order to attack the hierarchy
problem by adopting a theory space. We showed one-loop vacuum energies of
such models can be evaluated by using the determinant of the mass matrices
obtained by our method stated in earlier sections.

We have seen that there are not always genuine continuum limits in large
$N$ for general theory spaces.
In Sec.~\ref{sec7}, we have also found that contributions of
$V_0$ expressed in logarithmic functions remain in general. They can be
compensated by addition of bosonic or fermionic free fields with
appropriate mass in some cases.%
\footnote{In order to apply the models to the hierarchy problem, there
must be other matter fields coupled to the fields in the theory space.
Therefore, it may not be a great difficulty in the model.}

In future work, we wish to study one-loop energy density in models of
deconstructed warped (theory) space \cite{FK,AKMY,KS,CEG,RDT,BFPP}.
Although it is difficult to evaluate the determinants in a closed form in
such a model, calculation based on recurrence relations would be
suitable for a computer.
It is also interesting to investigate the recent model of deconstruction
of torus with magnetic flux \cite{TT}.%
\footnote{A partially deconstructed model with flux has been
considered more than a decade ago \cite{CKS}.}

If we would like to deal with the matrices related with more complicated
graphs or higher dimensional lattices, we confront other difficulties.%
\footnote{Even in the case with differential operators in higer
dimensions, there is a problem of degeneracy, which becomes an origin of
another divergence \cite{DK}.}
The graph Laplacians of generic graphs cannot be expressed by 
tridiagonal matrices.
Though, fortunately, it is known that arbitrary square matrices can be
systematically tridiagonalized by the Householder method
\cite{Householder} (see also Refs.~\cite{Lanczos, Givens}).
Thus, in principle, our Gel'fand--Yaglom-type method can be applied to
the matrix with the general graph structure.

Finally, we add a comment on exclusion of zero modes.
Zero modes of operators which appear in quantum field theory have crucial
meanings related with nonperturbative aspects of the theory (see for
example, the first section of Ref.~\cite{FFG}).
In our present paper, we considered mass terms in almost all examples
and the cases with zero modes can be considered as the limit that the
value of mass goes to zero. Because we considered the vacuum energies and
their dependence on the parameters in this paper, the analysis is
just sound. Moreover, it is known that, if a matrix is expressed as a
graph Laplacian of a simple graph (as in each example in this paper), the
matrix has a single zero modes. Therefore, further analysis on zero modes,
if necessary, could be fulfilled appropriately.

\section*{Data availability}
No data were used to support this study.

\section*{Conflicts of interest}
The authors declare that there are no conflicts of interest regarding the
publication of this paper.



\end{document}